\documentclass[twocolumn]{aastex701}
\usepackage{graphicx} 
\usepackage{natbib}
\usepackage{gensymb} 
\usepackage{appendix}
\usepackage[T1]{fontenc}
\usepackage{caption}
\usepackage{subcaption}
\usepackage{dblfloatfix}
\usepackage{tabularx}
\usepackage{tabularray}
\usepackage{float}
\usepackage[final]{microtype}
\usepackage{amsmath}
\begin{document}

\title{Chaotic Dynamics and Zero-Velocity Structures in the Pluto--Charon CR3BP}

\author{Abdul Wahab Jbara}
\affiliation{Independent Researcher}
\email[show]{aboudjbara.research@gmail.com}

\date{2025}

\begin{abstract}

Pluto and Charon are a dwarf binary system with a high mass ratio ($\mu \approx 0.109$), preventing the existence of stable Trojan companions. This instability creates ideal intersections for low-energy pathways that spacecraft can traverse and serves as a test case for stability at Lagrange points in high-$\mu$ binaries. This study models the Pluto–Charon system in the planar Circular Restricted Three-Body Problem (CR3BP) to validate the theoretical instability of its Lagrange points. Using fourth-order Runge–Kutta integration, we compute zero-velocity curves, Lagrange point locations, and compare tadpole and horseshoe trajectories of a massless particle with those of low-$\mu$ systems. The simulations reveal chaotic and unpredictable motion, where small variations in initial parameters lead to drastically different outcomes. These findings numerically confirm that Pluto–Charon cannot host long-lived Trojan companions, demonstrating the computational and educational value of CR3BP models to display chaotic transport in binary systems. Future extensions could incorporate the spatial or elliptic restricted three-body problem to further refine the analysis of instability in high-$\mu$ binaries.

\end{abstract}

\section{Introduction}
\label{sec:introduction}
The Circular Restricted Three-Body problem (CR3BP) is a classic problem of celestial mechanics
and is used to model two primaries and a third, significantly smaller object, used to show the dynamical effects of the primaries' gravitational force on the third body, usually a particle or a spacecraft; since the third body has no practical mass, it exerts no gravitational influence on the other two bodies \citep{1969SvA....13..364S}. This study aimed to observe the extent of the chaotic trajectories a massless particle could take under the influence of Pluto and Charon, which are considered unstable due to their high mass ratio $(8:1)$. The analysis was confined to the orbital plane, considering that the Pluto--Charon system has an inclination ($i$) $< 1\degree$, as well as for simpler visualization, therefore, the planar approximation is justified \citep{doi:10.1126/science.aad1815}, \citep{Buie_2006}. Pluto and Charon's high mass ratio, in particular, creates strong instability in contrast to systems such as Sun--Earth, which has a mass ratio of $\sim333000:1$, or Sun--Jupiter, which has a mass ratio of $\sim1000:1$, both of which are far larger than Pluto--Charon \citep{1966JRASC..60..167C}. This ratio is high enough that Pluto and Charon are considered a binary system, and are usually called a double dwarf planet.

The CR3BP includes five Lagrange points, each of which is considered a point of equilibrium under stable conditions. These points lie around the primaries and are the locations where the combined gravitational and centrifugal forces of both bodies balance out. Euler discovered the first three Lagrange points. Lagrange later identified the remaining two Lagrange points, completing the set of equilibrium points; these last two points, known as $L_4$ and $L_5$, form an equilateral triangle with the primaries and are regions where asteroids would naturally accumulate and remain in stable equilibrium for extended durations. It is important to locate these Lagrange points to identify the position of invariant manifolds, often called `tunnels', which are low-energy pathways in space that make it easier for a spacecraft to pass through, requiring less fuel and overall energy \citep{Calleja_2012} \citep{articl2e}. However, in the case of Pluto--Charon, the Lagrange points are all unstable, and any objects at these points won't be able to maintain their position for a lengthy duration of time.

The CR3BP also uses a rotating frame, first introduced by Euler, in which both bodies rotate around a common point, a center of mass, known as the barycenter. Using a rotating frame is convenient not only for its simplicity, but also because of how the CR3BP will be visually represented, with the primaries remaining seemingly fixed in position. This rotating frame introduces the Jacobi constant ($C$), which is a constant of motion, and constrains possible trajectories into allowed and forbidden regions (Zero Velocity Curves) (see Section \ref{subsec:zerovelocityfunction}). 

The case of Pluto and Charon is particularly interesting due to the degree of instability, where even small alterations can transform trajectories into drastically divergent and chaotic outcomes. Today, the CR3BP is utilized in trajectory design for complex spacecraft, especially for missions to Lagrange points, and the model is implemented directly in software such as Ansys Systems Tool Kit (STK). In this study, the planar CR3BP is applied to the Pluto–Charon system to numerically demonstrate and validate its theoretical instability and the chaotic evolution of trajectories near its Lagrange points, emphasizing the computational and educational value of such simulations.

\section{Methodology}
\label{sec:methodology}
\subsection{Positions of the Primaries}
\label{subsec:positionsofprimaries}
The orbit of Charon around Pluto is nearly circular, with an eccentricity of $\sim 0.005$, as well as a low inclination, so restricting the system to the orbital plane introduces negligible error. Therefore, the primaries and the massless particle will be modeled on an $x$-$y$ coordinate frame \citep{RHODEN201511}.

Following the standard CR3BP normalization, we set the distance between the primaries, the total mass of the two primaries, and the angular velocity to 1. We will then calculate the mass ratio between the primaries using:
\begin{equation}
    \mu = \frac{m_{Charon}}{m_{Pluto} + m_{Charon}}
\end{equation}
The masses of Pluto and Charon are $\sim 1.3\times10^{22}$ and $1.6\times10^{21}$, respectively \citep{article}. From the previous equation, we get:
\begin{equation}
    \mu \approx 0.109
\end{equation}
With the normalized distance between Pluto and Charon, and their barycenter at the point $(0,0)$, Pluto is placed at $(-\mu,0)$ and Charon at $(1-\mu,0)$. This forms the core of the CR3BP, and these conventions will be used in subsequent figures.

\subsection{Effective Potential $\Omega$}
\label{subsec:effectivepotential}
A central quantity of the CR3BP is the effective potential ($\Omega$) at each point in the rotating frame. The frame may be seen as a numerical grid, each point associated with a value of $\Omega$. $\Omega$ is defined as the combined potential of the gravitational forces of both primaries with the centrifugal force of the rotating frame. It is given by the following equation:

\begin{equation}
    \Omega(x,y) = \frac{1}{2}(x^2 + y^2) + \frac{1-\mu}{r_1} + \frac{\mu}{r_2}
\end{equation}

where $r_1$ and $r_2$ denote the distances from Pluto and Charon, respectively. (See Appendix \ref{sec:appendix_distances}).

\subsection{Zero-Velocity Function and the Jacobi Constant}
\label{subsec:zerovelocityfunction}
Once the massless particle is given a velocity at a certain point, the forbidden and allowed regions are fixed by the Jacobi constant ($C$). These regions are separated by Zero-Velocity Curves (ZVCs). It is possible to determine whether a region permits movement by the relation between $\Omega$ and the particle's $C$. The zero-velocity function is defined as:
\begin{equation}
    Z(x,y) = 2\Omega(x,y)
\end{equation}
Therefore, each point on the $Z(x,y)$ grid is simply twice the corresponding value of the point on the $\Omega(x,y)$ grid. 
The Jacobi Constant, $C$, is defined by
\begin{equation}
    C=Z(x,y) - (\dot{x}^2 + \dot{y}^2)
\end{equation}
It is important to note that $C$ is conserved throughout all movement (see Appendix \ref{sec:appendix_jacobi}). This conservation is expressed through the change in velocity as $Z(x,y)$ also changes. $C$ is the only conserved quantity in the planar CR3BP, and since the system has two degrees of freedom $(x,y,\dot{x}, \dot{y})$, which is more than one constant can control, motion can diverge from clean results, leading to chaotic behavior.

The value of $C$ depends on both position and velocity. Lower $C$ corresponds to higher energy and fewer forbidden regions, while a higher $C$ corresponds to fewer allowed regions and more restricted motion. When $\dot{x} = \dot{y} = 0$, $C$ takes its maximum value and will be equal to $2\Omega(x,y)$. This means that any points where $C=Z(x,y)$ will be areas where the massless particle has zero velocity, and going beyond that region will require negative energy, which is impossible. Hence, the locus of points where $C=Z(x,y)$ will form the zero-velocity curves (ZVCs). Regions where $C<Z(x,y)$ are forbidden, while motion is allowed only where $C\ge Z(x,y)$.

\subsubsection{Zero Velocity Curves}
\label{subsubsec:ZVC}
The areas with the lowest value of $C$ are $L_4$ and $L_5$, located above and below the primaries, respectively. They represent two Lagrange points of the system, areas of equilibrium, where the combined forces balance out. The first zero-velocity contours begin in this region.
\begin{figure}[h]
    \centering
    \includegraphics[width=1.1\linewidth]{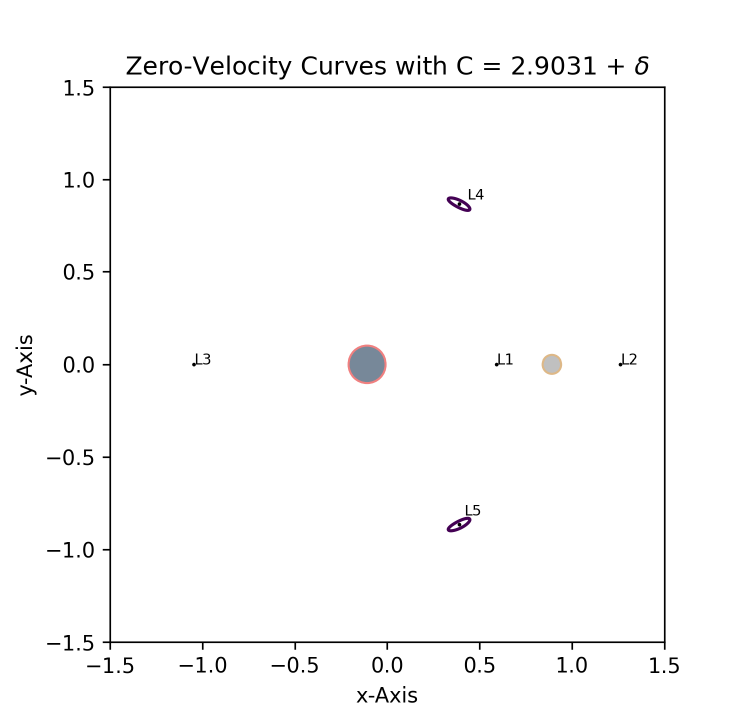}
    \caption{Purple contours mark the level set $C=2.9031+\delta$, with $\delta = 1\times10^{-3}$. Pluto (dark gray) is at $(-\mu,0)$ and Charon (light gray) is at $(1-\mu,0)$}
    \label{fig:F1PC}
\end{figure}

The five Lagrange points are denoted $L_1$-$L_5$. It can be seen in Figure \ref{fig:F1PC} that the ZVCs originate around $L_4$ and $L_5$, and as the value of $C$ increases, the ZVCs will expand to encompass the remaining Lagrange points. Movement is forbidden for the massless particle in this $C$ within those regions; motion is allowed in the remaining area. The value of $C\approx2.903$ is calculated with $\dot{x}=\dot{y}=0$, reducing the Jacobi constant equation to:
\begin{equation}
    C=2\Omega(x,y)
\end{equation}

A small offset $\delta$ is added to the value of $C$; without it, the ZVC at $C_{L_4}$ would collapse to a single point, rendering it invisible.

\begin{figure}[htbp]
        \centering 

        \begin{subfigure}[t]{0.495\textwidth} 
            \centering
            \includegraphics[width=1.1\linewidth]{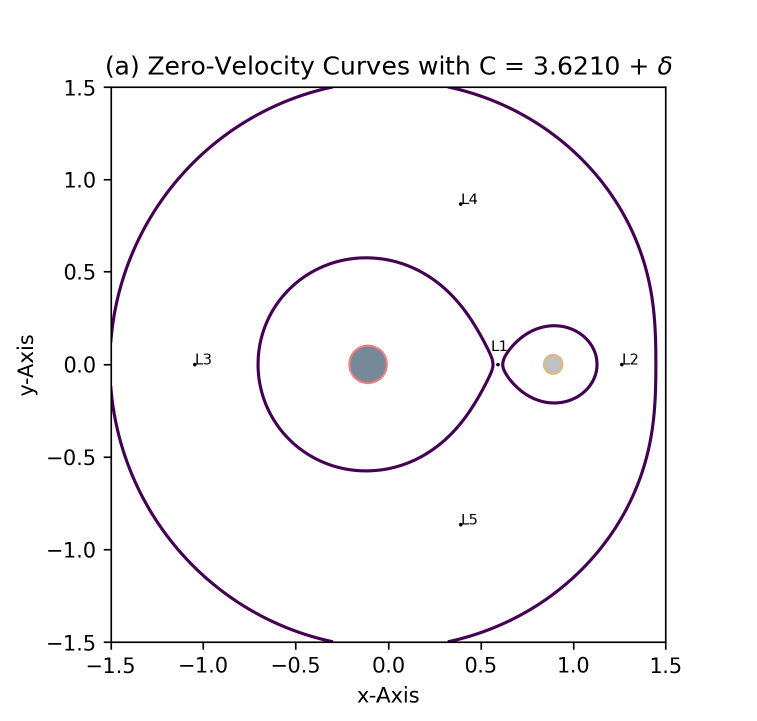} 
            \caption{ZVC for $C=3.6210+\delta$, with $\delta$ = $1\times10^{-2}$. $L_1$ neck is closed; particle cannot pass through. Pluto (dark gray) is at $(-\mu,0)$ and Charon (light gray) is at $(1-\mu,0)$}
            \label{fig:F2aPC}
        \end{subfigure}
        \hfill 
        \begin{subfigure}[t]{0.495\textwidth}
            \centering
            \includegraphics[width=1.1\linewidth]{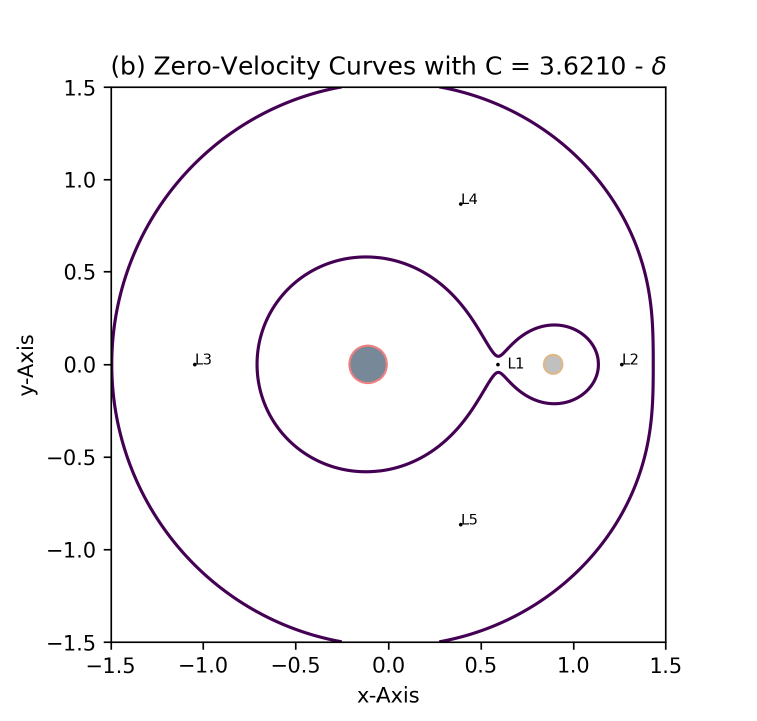} 
            \caption{ZVC for $C=3.6210+\delta$, with $\delta =-1\times10^{-2}$. $L_1$ neck is open, allowing passage. Pluto (dark gray) is at $(-\mu,0)$ and Charon (light gray) is at $(1-\mu,0)$}
            \label{fig:F2bPC}
        \end{subfigure}
        \caption{Figures show $C=3.6210+ \delta$, with $\delta=\pm1\times10^{-2}$. All $L$ points are in the forbidden region in $(a)$; A difference of only $2\delta$ in $C$ can alter connectivity between regions.}
        \label{fig:two_column_figure}
\end{figure}

Since the forbidden region expands as the value of $C$ increases, you can notice how certain areas will open and close based on small differences in $C$ (see Figure \ref{fig:two_column_figure})

The sole difference between Figures \ref{fig:F2aPC} and \ref{fig:F2bPC} is $2\delta$ in $C$, a small difference in energy, which alone causes drastic differences in trajectory, opening new pathways for the particle. This introduces the chaotic nature of the Pluto--Charon system, and demonstrates its extreme sensitivity to small parameter changes.

\subsection{Locating Lagrange Points}
\label{sub:locatinglagrangepoints}
\subsubsection{Triangular Lagrange Points}
\label{subsub:triangularlagrangepoints}
$L_4$ and $L_5$, the triangular Lagrange points, are located at the vertices of an equilateral triangle formed with the two primaries. This simplifies locating them; $L_4$ can be found using the following equation:
\begin{equation}
    L_4=\left(\frac{1}{2}- \mu,\frac{\sqrt{3}}{2}\right)
\end{equation}
And $L_5$ can be found with the corresponding equation:
\begin{equation}
    L_5=\left(\frac{1}{2}- \mu,-\frac{\sqrt{3}}{2}\right)
\end{equation}
These two points are symmetric with respect to the $x$-axis.
\subsubsection{Collinear Lagrange Points}
\label{subsub:collinearlagrangepoints}
These three points, $L_1$, $L_2$, and $L_3$, lie along the $x$-axis on the line connecting the primaries, and cannot be expressed in closed form, so the Newton-Raphson iteration was applied to locate the roots (see Appendix \ref{lpointappendix}).
\subsubsection{Lagrange Points on Figures}
\label{subsubsec:lagrangepointsonfigures}
In the previous figures, the chosen values of $C$ were taken directly from the Lagrange points. For example, in Figure \ref{fig:F1PC} we use $C=2.9031+\delta$, where $2.9031$ is the exact value of $C$ at $L_4$ and $L_5$ when $\dot{x}=\dot{y}=0$. Since Lagrange points are points of equilibrium, the zero velocity curve passes exactly through $L_4$ and $L_5$. The same applies to figure \ref{fig:two_column_figure}. If $C=3.6210$, then the zero-velocity curve would pass exactly through $L_1$.

\subsection{Updating the Particle's State}
\label{subsec:updatingpositionofparticle}
Since the particle's position changes, we need to update the $x$, $y$, $\dot{x}$, and $\dot{y}$ values. Because $\Omega$ varies in the rotating frame, the acceleration depends on position. As a result, simple Euler updates will be inaccurate unless the time step is sufficiently small.
\subsubsection{Equations of Motion}
\label{subsubsec:equationsofmotion}
The planar equations of motion in the rotating frame are: \citep{1999ssd..book.....M}
\begin{equation}
    \ddot{x} = \frac{\partial \Omega}{\partial x}+2\dot{y}
\end{equation}
\begin{equation}
    \ddot{y} = \frac{\partial \Omega}{\partial y}-2\dot{x}
\end{equation}

\subsubsection{Runge-Kutta Fourth Order Method}
\label{subsubsec:rk4}
To integrate the equations of motion, we employed the Runge-Kutta Fourth Order method (RK4). This method determines the slopes across four time steps and creates a weighted average, yielding fourth-order accuracy. Since the acceleration varies significantly at different positions on the rotating frame, RK4 provides far greater accuracy and stability than simpler methods such as Euler integration. 
\begin{equation}
    y_\text{new} = y_0+\frac{h}{6}(k_1+2k_2+2k_3+k_4)
\end{equation}
Where $h$ is the time step, $y$ the state vector $(x,y,\dot{x},\dot{y})$, and $k_{1-4}$ denote slope evaluations used to form the weighted average (see Appendix \ref{appendix:rk4})
Each trajectory simulation will be integrated for a time duration $T$. The number of RK4 repetitions is therefore $R = \frac{T}{h}$. It's also important to note that after each RK4 step, we recalculate $C$ and compare it with its initial value; conservation of $C$ was a key indicator of the success of RK4 without error.

\subsection{Canonical Tadpole and Horseshoe Orbits}
\label{subsec:canonicaltadpoleandhorseshoeorbits}
Tadpole and horseshoe orbits are well-known periodic solutions of the CR3BP. They exist when $L_4$ and $L_5$ are stable, which requires a sufficiently small mass ratio $\mu$. Tadpoles correspond to librations around $L_4$ or $L_5$, and horseshoes correspond to extended librations spanning $L_4$, $L_3$, and $L_5$ in an indefinite loop. To reproduce these orbits, the value of $\mu$ must satisfy $\mu \le0.03852$, where $0.03852$ represents Routh's critical mass ratio \citep{SOKOLSKII1975342}; smaller values of $\mu$ yield more coherent librations.
\begin{figure}[h]
    \centering
    \includegraphics[width=1.1\linewidth]{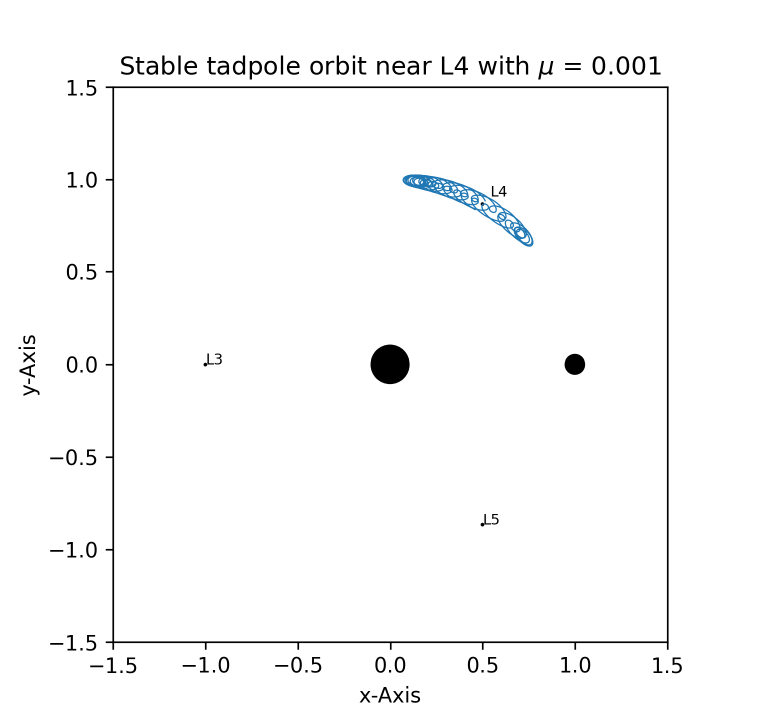}
    \caption{Example of a tadpole orbit librating around $L_4$. The primaries are shown as black circles. Integration parameters: $h=5\times 10^{-4}$, $C_{\text{target}}=C_{L_4}+\delta$, with $\delta=-5\times 10^{-4}$ tangential initial velocity, starting coordinates $(L_{4_x}-0.04,L_{4_y}-0.03)$. Drift in $C$: $\sim 6.7\times 10^{-15}$. The trajectory is plotted in blue.}
    \label{fig:F5PC}
\end{figure}

Figure \ref{fig:F5PC} shows a classic tadpole orbit, where a Trojan, a small body, librates around $L_4$ (or around $L_5$ in symmetry) indefinitely. Compared to horseshoe orbits, tadpoles have smaller amplitudes but share the same orbital period as the secondary (see Figure \ref{fig:F4PC}).

\subsubsection{Initial Velocity Parameters}
\label{subsubsec:initialvelocityparameters}
Numerical simulations employ a $C_{target}$, set manually, from which the corresponding $|v|$ is calculated (see Appendix \ref{appendix:targetjacobiconstant}). In the case of Figure \ref{fig:F5PC}, the $C_\text{target}$ was $C_{L_4}-\delta$ where $\delta$ is $5\times 10^{-4}$.

\begin{figure}[h]
    \centering
    \includegraphics[width=1.1\linewidth]{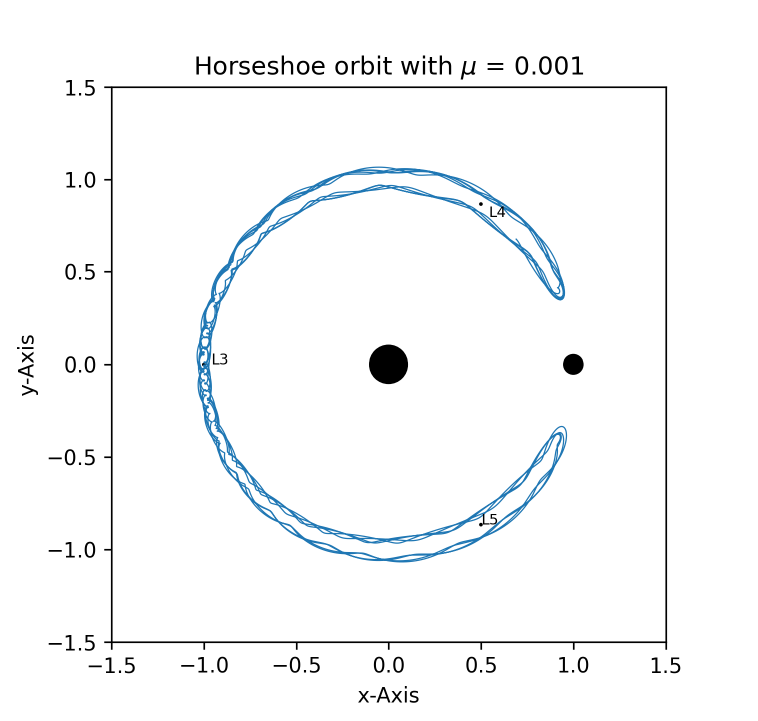}
    \caption{Example of a horseshoe orbit. The primaries are shown as black circles. Integration parameters: $h=5\times 10^{-4}$, $C_{\text{target}}=C_{L_3}+\delta$, with $\delta=-5\times10^{-5}$, vertical initial velocity, starting coordinates $(L_{3_x}-0.027,L_{3_y})$. Drift in $C$: $\sim 4.9\times 10^{-15}$. The trajectory is plotted in blue.}
    \label{fig:F4PC}
\end{figure}

For numerical simulations, the initial direction can have a significant influence on the final trajectory. For example, in Figure \ref{fig:F5PC}, a tangential $v$ was used, directed tangentially clockwise about the barycenter at $(0,0)$. In other cases, it may be preferable to directly specify the $\dot{x}$ and $\dot{y}$ components (see Figure \ref{fig:F4PC}).

\subsubsection{Collisions}
\label{subsubsec:collisions}
Collisions with primaries may occur during integrations. In such instances, the numerical simulation is terminated, and the reported collision time $T$, in normalized units, corresponds to the collision point (see Figure \ref{fig:F3PC}). A collision is defined numerically as $r_1$ or $r_2 < 1\times10^{-6}$.

\section{Results}
\label{sec:results}
In this section, we present and compare the numerical results for the Pluto--Charon CR3BP. In Figures \ref{fig:F5PC} and \ref{fig:F4PC}, we had ideal results due to a small $\mu$, with smooth and predictable trajectories. Increasing $\mu$ fundamentally alters the system's stability, producing irregular and chaotic trajectories. 

\subsection{Tadpoles in the Pluto--Charon System}
\label{subsec:tadpolesinthePluto--Charonsystem}
\begin{figure}[h]
    \centering
    \includegraphics[width=1.1\linewidth]{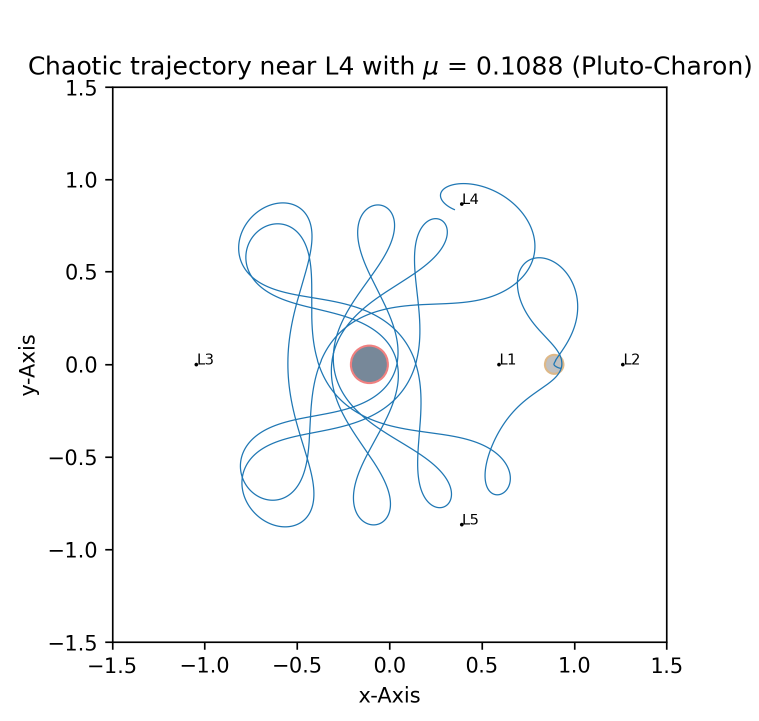}
    \caption{Attempt at tadpole orbit for Pluto--Charon. Pluto (dark gray) is at $(-\mu,0)$ and Charon (light gray) is at $(1-\mu,0)$. Integration parameters: $h=5\times 10^{-4}$, $C_{\text{target}}=C_{L_4}+\delta$, with $\delta=-5\times10^{-4}$, tangential initial velocity, starting coordinates $(L_{4_x}-0.04,L_{4_y}-0.03)$. Drift in $C$: $\sim 5.1\times 10^{-9}$. The trajectory is plotted in blue.}
    \label{fig:F3PC}
\end{figure}

Recall that a canonical tadpole librates around $L_4$ or $L_5$ at a low $\mu$ indefinitely. In Figure \ref{fig:F3PC}, the particle diverges from its orbit around $L_4$, demonstrating the linear instability in triangular Lagrange points at high mass ratios; the orbit is chaotic and escapes, hence Pluto--Charon cannot host long-lived Trojan companions.

Additionally, the particle collides with Charon near the end of the numerical integration. This occurs at $T \approx 37$; instead of an eventual chaotic trajectory, the tadpole attempt results in an impact, terminating the integration. 

\subsection{Horseshoes in the Pluto--Charon System}
\label{subsec:horseshoesinthePluto--Charonsystem}
Unlike tadpoles, horseshoes librate around both triangular Lagrange points ($L_4$ and $L_5$) by passing around $L_3$. Horseshoes occur when $C$ is slightly below $C_{L_3}$, such that the neck at $L_3$ is marginally open.

\begin{figure}[h]
    \centering
    \includegraphics[width=1.1\linewidth]{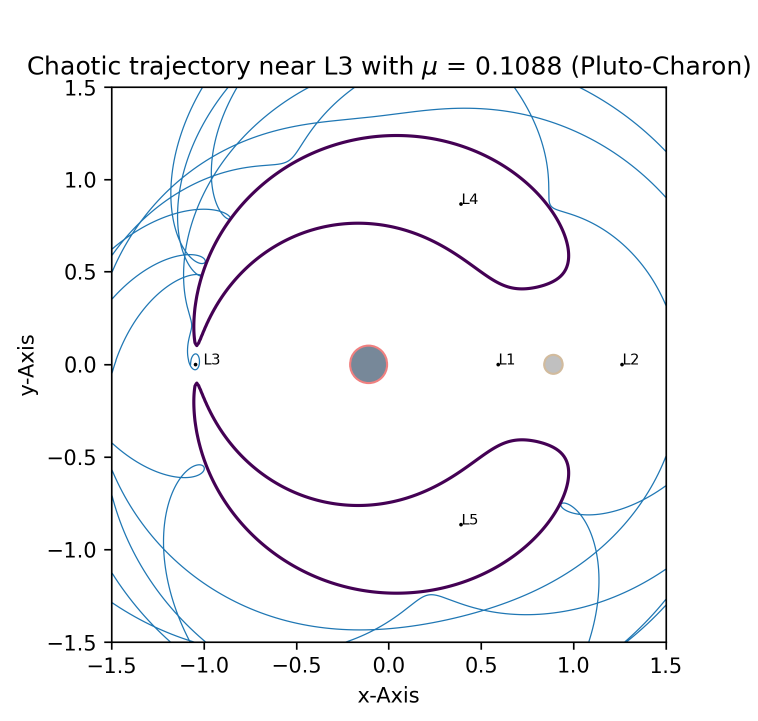}
    \caption{Attempt at horseshoe orbit for Pluto--Charon. Pluto (dark gray) is at $(-\mu,0)$ and Charon (light gray) is at $(1-\mu,0)$. Integration parameters: $h=5\times 10^{-4}$, $C_{\text{target}}=C_{L_3}+\delta$, with $\delta=-1\times10^{-3}$, vertical initial velocity, starting coordinates $(L_{3_x}-0.027,L_{3_y})$. Drift in $C$: $\sim 8.9\times 10^{-13}$. The trajectory is plotted in blue; purple contours show ZVCs.}
    \label{fig:F7PC}
\end{figure}

In Figure \ref{fig:F7PC}, the trajectory diverges after $\sim 2$ librations from the ideal horseshoe symmetry seen in Figure \ref{fig:F4PC} and becomes chaotic. The value of $\delta$ differs between Figures \ref{fig:F7PC} and \ref{fig:F4PC}. Although both are below the $C_{L_3}$ threshold and produce qualitatively similar behavior, we present the larger offset here because it makes the divergence visually more pronounced. The zero-velocity curves extend more widely in Figure \ref{fig:F7PC} than the thinner ZVCs in Figure \ref{fig:F4PC} due to Charon's substantial mass ratio and consequently larger forbidden regions for the same parameters.

\subsection{$L_1$ Neck Sensitivity}
\label{subsec:l1necksensitivity}
From Figure \ref{fig:two_column_figure}, the ZVCs at the $L_1$ neck opened and closed with a difference of $2\delta$. This directly affects the trajectory of the particle, allowing it to pass into separate regions through the $L_1$ neck. 

\begin{figure}[H]
        \centering 
        \begin{subfigure}[t]{0.495\textwidth} 
            \centering
            \includegraphics[width=1.1\linewidth]{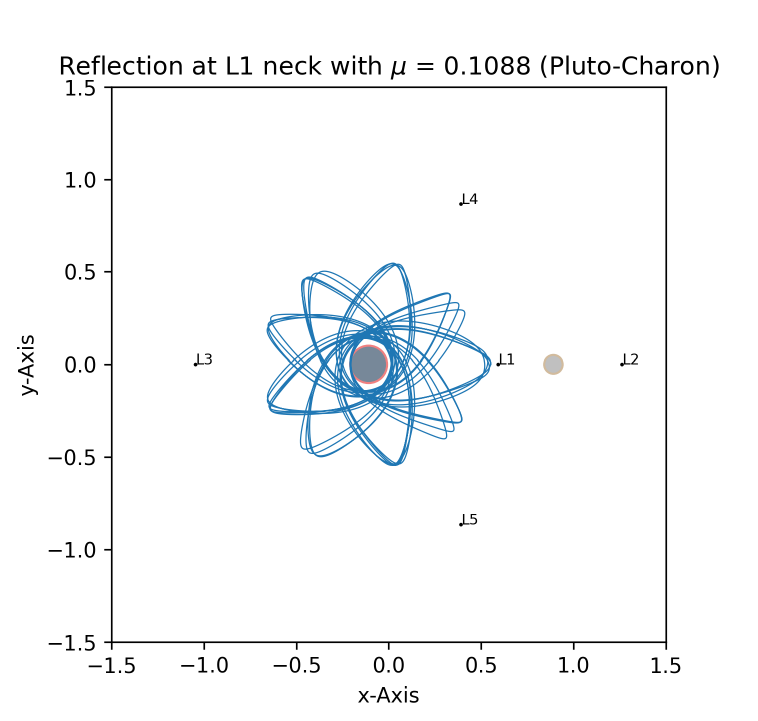} 
            \caption{Trajectory near Pluto with neck closed. Integration parameters: $h=5\times 10^{-4}$, $C_{\text{target}}=C_{L_1}+\delta$, with $\delta=1\times10^{-2}$, vertical initial velocity, starting coordinates $(L_{1_x}-0.04,L_{1_y})$. Drift in $C$: $\sim 4.9\times 10^{-9}$.}
            \label{fig:F6aPC}
        \end{subfigure}
        \hfill 
        \begin{subfigure}[t]{0.495\textwidth}
            \centering
            \includegraphics[width=1.1\linewidth]{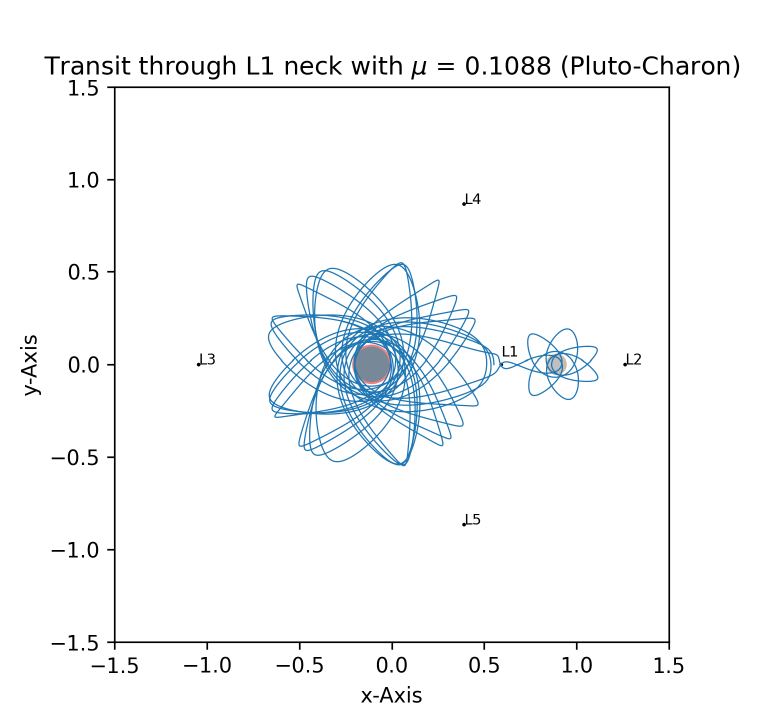} 
            \caption{Trajectory near Pluto with neck open. Integration parameters: $h=5\times 10^{-4}$, $C_{\text{target}}=C_{L_1}+\delta$, with $\delta=-1\times10^{-2}$, vertical initial velocity, starting coordinates $(L_{1_x}-0.04,L_{1_y})$. Drift in $C$: $\sim 8.1\times 10^{-8}$.}
            \label{fig:F6bPC}
        \end{subfigure}
        \caption{Pluto (dark gray) is at $(-\mu,0)$ and Charon (light gray) is at $(1-\mu,0)$. The trajectory is plotted in blue. Figure \ref{fig:F6bPC} shows the particle escaping into Charon's surrounding regions through $L_1$.}
        \label{fig:second_two_column_figure}
\end{figure}

As shown in Figure \ref{fig:F6bPC}, near-threshold orbits linger before crossing the neck, consistent with a high value of $C$. In this case, $C$ near the neck is slightly below $C_{L_1}$, and to maintain the initial $C$, velocity is reduced and the particle lingers before either reflecting to Pluto or transiting towards Charon. If $\delta<0$, the particle will transit into the Charon lobe, and if $\delta>0$, it will reflect into the Pluto lobe. This asymmetry of outcome despite similar short-term motion is another key indicator of a chaotic system, where small differences yield substantially variable outcomes. 

The transit in Figure \ref{fig:F6bPC} occurs over a period of $T\approx27$-$29$. This corresponds to the lingering effect that occurs at high values of $C$; the high $T$ indicates reduced velocity, and therefore, there is a delayed passage compared to typical neck-crossing times.

Figures \ref{fig:F6aPC} and \ref{fig:F6bPC} show apparent order; nonetheless, they do not represent stable librations and are temporary, unstable directions of motion. They will eventually diverge into chaos. An asteroid or spacecraft will not be able to oscillate within the Pluto lobe indefinitely, and instability will push it through the neck or eject it.

\section{Discussion}
\subsection{Model Assumptions and Limitations}
\label{subsec:modelassumptionsandlimitations}
One major assumption in the CR3BP is to neglect the mass and influence of the moving object, which, in this study, corresponds to the massless particle. In most realistic cases, the third body may have a large mass that could affect the orbit of the primaries. Neglecting this factor could create inaccurate representations of realistic systems or neglect certain chaotic trajectories that may occur. This makes the CR3BP unsuitable for systems where there are mutual perturbations between the bodies. 

In this model, only two dimensions were considered, omitting the $z$-axis, which implies that vertical stability was neglected and the third equation of motion was omitted. Excluding the third dimension also excludes entire classes of three-dimensional orbits, such as halo and Lissajous orbits; halo orbits, in particular, are important for space telescopes, such as JWST \citep{inproceedings}. Consequently, a two-dimensional model is inadequate for realistic mission design.

Most celestial orbits are elliptical and are not circular, as assumed in the CR3BP. However, the Pluto--Charon system has a low eccentricity value (see Section \ref{subsec:positionsofprimaries}), so the resulting error is minimized. Despite this, over long timescales, a periodic forcing is introduced, and over the course of multiple librations, it could cause further divergence, in addition to the already chaotic nature of the Pluto--Charon system. A more accurate method is the ER3BP, which considers an elliptical orbit. This causes the system to become time dependent, and the Jacobi constant $C$ is no longer strictly conserved.

\subsection{Dynamical Implications for Pluto--Charon}
\label{subsec:dynamicalimplicationsforplutocharon}
Our results confirm that long-lived Trojans cannot exist in the Pluto--Charon system due to the high $\mu$, consistent with previous findings \citep{Sicardy2010}. Short-term captures are possible, where an object may librate near a Lagrange point once or twice before diverging (see Section \ref{sec:results}). These results generalize to other high-$\mu$ binary systems, displaying how increased mass ratios destabilize libration regions.

The Pluto--Charon instability prevents long-term orbits, though it also creates opportunities for low-energy transport between the Pluto and Charon lobes. The unstable manifolds of Pluto--Charon may overlap with manifolds from other systems, such as Sun--Pluto, allowing a spacecraft to cross from one to another with minimal propulsion. These interconnected manifolds are known as `interplanetary superhighways', and exploiting them can significantly reduce fuel expenditure for spacecraft.

\section{Conclusion}
\label{sec:conclusion}
In this work, we modeled the planar CR3BP for the Pluto–Charon system, with $\mu \approx 0.109$. We presented classical horseshoe and tadpole orbits for a low-$\mu$ system, then compared them with corresponding simulations in the Pluto–Charon case, which is intrinsically unstable. We also examined trajectories with the $L_1$ neck open and closed. The Pluto–Charon system exhibits strong sensitivity to initial conditions, where small changes in position or velocity lead to drastically different trajectories. These results numerically validate the theoretical instability of the triangular Lagrange points for high-$\mu$ binaries, confirming that Pluto–Charon cannot sustain long-lived Trojan companions. Future work could extend this analysis to the spatial or elliptic restricted three-body problem to further refine the understanding of chaotic transport and instability in high-$\mu$ systems.

\bibliographystyle{aasjournalv7}
\bibliography{references}

\begin{thebibliography}{}
\expandafter\ifx\csname natexlab\endcsname\relax\def\natexlab#1{#1}\fi
\providecommand{\url}[1]{\href{#1}{#1}}
\providecommand{\dodoi}[1]{doi:~\href{http://doi.org/#1}{\nolinkurl{#1}}}
\providecommand{\doeprint}[1]{\href{http://ascl.net/#1}{\nolinkurl{http://ascl.net/#1}}}
\providecommand{\doarXiv}[1]{\href{https://arxiv.org/abs/#1}{\nolinkurl{https://arxiv.org/abs/#1}}}

\bibitem[{M.~W. Buie {et~al.}(2006)Buie, Grundy, Young, Young, \& Stern}]{Buie_2006}
Buie, M.~W., Grundy, W.~M., Young, E.~F., Young, L.~A., \& Stern, S.~A. 2006, \bibinfo{title}{Orbits and Photometry of Pluto’s Satellites: Charon, S/2005 P1, and S/2005 P2,} The Astronomical Journal, 132, 290–298, \dodoi{10.1086/504422}

\bibitem[{R.~C. Calleja {et~al.}(2012)Calleja, Doedel, Humphries, Lemus-Rodríguez, \& Oldeman}]{Calleja_2012}
Calleja, R.~C., Doedel, E.~J., Humphries, A.~R., Lemus-Rodríguez, A., \& Oldeman, E.~B. 2012, \bibinfo{title}{Boundary-value problem formulations for computing invariant manifolds and connecting orbits in the circular restricted three body problem,} Celestial Mechanics and Dynamical Astronomy, 114, 77–106, \dodoi{10.1007/s10569-012-9434-y}

\bibitem[{G.~M. {Clemence}(1966){Clemence}}]{1966JRASC..60..167C}
{Clemence}, G.~M. 1966, \bibinfo{title}{{The Scale of the Solar System},} \jrasc, 60, 167

\bibitem[{G. Gomez {et~al.}(2001)Gomez, Koon, Lo, Marsden, Masdemont, \& Ross}]{articl2e}
Gomez, G., Koon, W., Lo, M., {et~al.} 2001, \bibinfo{title}{Invariant Manifolds, the Spatial Three-Body Problem and Space Mission Design,} Advances in the Astronautical Sciences, 109

\bibitem[{P. Llanos \& A. Matas(2022)Llanos \& Matas}]{inproceedings}
Llanos, P., \& Matas, A. 2022, \bibinfo{title}{AAS 22-623 JAMES WEBB SPACE TELESCOPE TRAJECTORY, COMMUNICATIONS, AND INSTRUMENTATION COMPREHENSIVE ANALYSIS,}

\bibitem[{C.~D. {Murray} \& S.~F. {Dermott}(1999){Murray} \& {Dermott}}]{1999ssd..book.....M}
{Murray}, C.~D., \& {Dermott}, S.~F. 1999, {Solar System Dynamics}, \dodoi{10.1017/CBO9781139174817}

\bibitem[{A.~R. Rhoden {et~al.}(2015)Rhoden, Henning, Hurford, \& Hamilton}]{RHODEN201511}
Rhoden, A.~R., Henning, W., Hurford, T.~A., \& Hamilton, D.~P. 2015, \bibinfo{title}{The interior and orbital evolution of Charon as preserved in its geologic record,} Icarus, 246, 11, \dodoi{https://doi.org/10.1016/j.icarus.2014.04.030}

\bibitem[{B. Sicardy(2010)Sicardy}]{Sicardy2010}
Sicardy, B. 2010, \bibinfo{title}{Stability of the triangular Lagrange points beyond Gascheau’s value,} Celestial Mechanics and Dynamical Astronomy, 107, 145, \dodoi{10.1007/s10569-010-9259-5}

\bibitem[{A. Sokol'skii(1975)Sokol'skii}]{SOKOLSKII1975342}
Sokol'skii, A. 1975, \bibinfo{title}{Stability of the lagrange solutions of the restricted three-body problem for the critical ratio of the masses: PMM vol. 39, n≗2, 1975, pp. 366–369,} Journal of Applied Mathematics and Mechanics, 39, 342, \dodoi{https://doi.org/10.1016/0021-8928(75)90158-6}

\bibitem[{S. Stern(2003)Stern}]{article}
Stern, S. 2003, \bibinfo{title}{The Pluto-Charon Systems,} Annual Review of Astronomy and Astrophysics, 30, 185, \dodoi{10.1146/annurev.aa.30.090192.001153}

\bibitem[{S.~A. Stern {et~al.}(2015)Stern, Bagenal, Ennico, Gladstone, Grundy, McKinnon, Moore, Olkin, Spencer, Weaver, Young, Andert, Andrews, Banks, Bauer, Bauman, Barnouin, Bedini, Beisser, Beyer, Bhaskaran, Binzel, Birath, Bird, Bogan, Bowman, Bray, Brozovic, Bryan, Buckley, Buie, Buratti, Bushman, Calloway, Carcich, Cheng, Conard, Conrad, Cook, Cruikshank, Custodio, Ore, Deboy, Dischner, Dumont, Earle, Elliott, Ercol, Ernst, Finley, Flanigan, Fountain, Freeze, Greathouse, Green, Guo, Hahn, Hamilton, Hamilton, Hanley, Harch, Hart, Hersman, Hill, Hill, Hinson, Holdridge, Horanyi, Howard, Howett, Jackman, Jacobson, Jennings, Kammer, Kang, Kaufmann, Kollmann, Krimigis, Kusnierkiewicz, Lauer, Lee, Lindstrom, Linscott, Lisse, Lunsford, Mallder, Martin, McComas, McNutt, Mehoke, Mehoke, Melin, Mutchler, Nelson, Nimmo, Nunez, Ocampo, Owen, Paetzold, Page, Parker, Parker, Pelletier, Peterson, Pinkine, Piquette, Porter, Protopapa, Redfern, Reitsema, Reuter, Roberts, Robbins, Rogers, Rose, Runyon, Retherford,
  Ryschkewitsch, Schenk, Schindhelm, Sepan, Showalter, Singer, Soluri, Stanbridge, Steffl, Strobel, Stryk, Summers, Szalay, Tapley, Taylor, Taylor, Throop, Tsang, Tyler, Umurhan, Verbiscer, Versteeg, Vincent, Webbert, Weidner, Weigle, White, Whittenburg, Williams, Williams, Williams, Woods, Zangari, \& Zirnstein}]{doi:10.1126/science.aad1815}
Stern, S.~A., Bagenal, F., Ennico, K., {et~al.} 2015, \bibinfo{title}{The Pluto system: Initial results from its exploration by New Horizons,} Science, 350, aad1815, \dodoi{10.1126/science.aad1815}

\bibitem[{V. {Szebehely} \& E. {Grebenikov}(1969){Szebehely} \& {Grebenikov}}]{1969SvA....13..364S}
{Szebehely}, V., \& {Grebenikov}, E. 1969, \bibinfo{title}{{Theory of Orbits-The Restricted Problem of Three Bodies.},} \sovast, 13, 364

\end{thebibliography}

\appendix
\section{Distances from Pluto and Charon}
\label{sec:appendix_distances}
$r_1$ and $r_2$ are the distances from both primaries, and were calculated using the distance equation:
\begin{equation}
    r = \sqrt{(x_1-x_0)^2 + (y_1-y_0)^2}
\end{equation}
where $x_0, y_0$ represent the $x$ and $y$ coordinates of the primary, and $x_1,y_1$ represent the $x$ and $y$ coordinates of the current point. Pluto and Charon have coordinates $(-\mu,0)$ and $(1-\mu,0)$, respectively; therefore, $r_1$ and $r_2$ can be given using the following equations:
\begin{equation} 
    r_1 = \sqrt{(x+\mu)^2 + (y)^2}
\end{equation}
\begin{equation}
    r_2 = \sqrt{(x - 1 +\mu)^2 + (y)^2}
\end{equation}
where $r_1$ is the distance of the current point from Pluto, and $r_2$ is the distance of the current point from Charon.

\section{Origin of the Jacobi Constant}
\label{sec:appendix_jacobi}
Starting from the planar equations of motion from Section \ref{subsubsec:equationsofmotion}:
\begin{equation}
    \ddot{x} -2\dot{y} = \frac{\partial \Omega}{\partial x}
\end{equation}
\begin{equation}
    \ddot{y} +2\dot{x} = \frac{\partial \Omega}{\partial y}
\end{equation}
Next, we multiply the $x$ equation by $\dot{x}$ and the $y$ equation by $\dot{y}$, and add both equations together:
\begin{equation}
    \dot{x}\ddot{x} -2\dot{x}\dot{y} + \dot{y}\ddot{y} + 2\dot{x}\dot{y}= \frac{\partial \Omega}{\partial x}\dot{x} + \frac{\partial \Omega}{\partial y}\dot{y}
\end{equation}
Terms cancel out:
\begin{equation} \label{eq:jacobib7}
    \dot{x}\ddot{x} + \dot{y}\ddot{y}= \frac{\partial \Omega}{\partial x}\dot{x} + \frac{\partial \Omega}{\partial y}\dot{y}
\end{equation}

Isolating the left side, we recognize the derivative:
\begin{equation}
    \frac{d}{dt}\left(\frac{1}{2}(\dot{x}^2+\dot{y}^2)\right)
\end{equation}
And expanding the brackets:
\begin{equation}
    \frac{d}{dt}\left(\frac{1}{2}\dot{x}^2 \right)+\frac{d}{dt}\left(\frac{1}{2}\dot{y}^2 \right)
\end{equation}
By the chain rule, $\frac{d}{dt}\frac{1}{2}\dot{x}^2 = \dot{x}\ddot{x}$, and similarly for $y$.
\begin{equation}
    \frac{d}{dt}\left(\frac{1}{2}(\dot{x}^2+\dot{y}^2)\right) = \dot{x}\ddot{x} + \dot{y}\ddot{y}
\end{equation}
Next, consider the right-hand side of equation Eq.~(\ref{eq:jacobib7}) and expand:
\begin{equation}
    \frac{\partial \Omega}{\partial x}\dot{x} + \frac{\partial \Omega}{\partial y}\dot{y} =\frac{\partial \Omega}{\partial x}\frac{dx}{dt} + \frac{\partial \Omega}{\partial y}\frac{dy}{dt}
\end{equation}
Applying the multivariable chain rule, we can shorten the expression:
\begin{equation}
  \frac{\partial \Omega}{\partial x}\frac{dx}{dt} + \frac{\partial \Omega}{\partial y}\frac{dy}{dt} = \frac{d\Omega}{dt}
\end{equation}
Rewriting Eq.~(\ref{eq:jacobib7}) with the new changed terms gives us:
\begin{equation}
    \frac{d}{dt}\left(\frac{1}{2}(\dot{x}^2+\dot{y}^2)\right) = \frac{d\Omega}{dt}
\end{equation}
Changing the subject gives us the following version of the equation:
\begin{equation} \label{eq:jacobib14}
    \frac{d}{dt}\left(\frac{1}{2}(\dot{x}^2+\dot{y}^2) - \Omega(x,y)\right) = 0
\end{equation}
This indicates that the bracketed expression remains constant over time, and that there are no total changes as the individual values inside the brackets change. We can turn this into a constant:
\begin{equation}
    E_J = \frac{1}{2}(\dot{x}^2+\dot{y}^2) - \Omega(x,y)
\end{equation}
And by convention, we multiply by -2 to simplify the expression and for easier use. This gives us the Jacobi Constant:
\begin{align}
    \boldsymbol{C = 2\Omega(x,y) - (\dot{x}^2+\dot{y}^2)}
\end{align}
where $\dot{x}$ and $\dot{y}$ represent the velocity of the massless particle. Because the time derivative of Eq.~(\ref{eq:jacobib14}) vanishes, $C$ remains constant over time, representing the conserved energy-like quantity in the rotating frame.

\section{Collinear Lagrange Point Equations}
\label{lpointappendix}
Taking the effective potential equation from Section \ref{subsec:effectivepotential} with $y=0$, because all the collinear points lie on the x-axis, we get the equation:
\begin{equation}
    \Omega(x,0)=\frac{1}{2}x^2 + \frac{1-\mu}{r_1}+\frac{\mu}{r_2}
\end{equation}
Since $y=0$, the equations for $r_1$ and $r_2$ have been simplified, and so $r_1=|x+\mu|$ and $r_2=|x-1+\mu|$. This gives the following equation:
\begin{equation}
    \Omega(x)=\frac{1}{2}x^2 + \frac{1-\mu}{|x+\mu|}+\frac{\mu}{|x-1+\mu|}
\end{equation}
The Lagrange points represent areas of stability, where any particle at exactly that point remains fixed. That means $\frac{\partial\Omega}{\partial x}=0$, so we differentiate the previous equation, which gives us the following (defined from now on as $f(x)$):
\begin{equation}
    f(x)=x-(1-\mu)\frac{x+\mu}{|x+\mu|^3}-\mu\frac{x-1+\mu}{|x-1+\mu|^3}
\end{equation}
Therefore, the equilibrium points will lie on values where $f(x)=0$. It's also possible to further simplify the equations for each particular Lagrange point, since they each lie in different regions.

The sign of each term depends on the region of $x$ relative to the primaries; therefore, the absolute values must be expanded to preserve the correct direction.

For example, point $L_1$ lies between the primaries on the positive x-axis, so the value of x for it will be positive. In the case of $\frac{x+\mu}{|x+\mu|^3}$, we can understand that $x+\mu$ will be a positive value between 0 and 1, so there is no need for the absolute sign in the denominator. For $\frac{x-1+\mu}{|x-1+\mu|^3}$, the numerator will be a negative value because we have already established that $x+\mu$ is a positive value between 0 and 1; therefore, $x-1+\mu$ must be a negative value between 0 and -1. This means that if we remove the absolute sign, the sign of the fraction will change; therefore, we must change the $-$ on the outside to a $+$ to maintain accuracy. All this gives us a final equation for $L_1$
\begin{equation}
    f_{L_1}(x) = x-\frac{1-\mu}{(x+\mu)^2}+\frac{\mu}{(x-1+\mu)^2}
\end{equation}
We can do the same for $L_2$ and $L_3$, obtaining:
\begin{equation}
    f_{L_2}(x) = x-\frac{1-\mu}{(x+\mu)^2}-\frac{\mu}{(x-1+\mu)^2}
\end{equation}
\begin{equation}
    f_{L_3}(x) = x+\frac{1-\mu}{(x+\mu)^2}+\frac{\mu}{(x-1+\mu)^2}
\end{equation}
To find their roots, we compute their derivatives, then apply Newton-Raphson iteration, using initial guesses for $x$.

\section{Runge-Kutta Fourth Order Method}
\label{appendix:rk4}
The Runge-Kutta Fourth Order (RK4) method works by evaluating the slope at four separate points, then combining them in a weighted average to achieve fourth-order accuracy.
We begin by taking the $x$ and $y$ coordinates of our starting point and getting the values for $\frac{\partial \Omega}{\partial x}$ and $\frac{\partial \Omega}{\partial y}$, by partially differentiating the effective potential equation. We then use the equations of motion from Section \ref{subsubsec:equationsofmotion} to determine the values of $\ddot{x}$ and $\ddot{y}$. Then we can define $k_1$ as the derivative of $(x,y,\dot{x}, \dot{y})$. This gives us the following:
\begin{equation}
    k_1=(\dot{x},\dot{y},\ddot{x},\ddot{y})
\end{equation}
Each $k_i$ will represent the above slope vector but at different points;
we can then use $k_1$ to get $k_2$ and so on:
\begin{equation}
    k_2=(x_n+\frac{h}{2}k_{1(x)},y_n+\frac{h}{2}k_{1(y)},\dot{x}_n+\frac{h}{2}k_{1(\dot{x})},\dot{y}_n+\frac{h}{2}k_{1(\dot{y})})
\end{equation}
$k_3$ is determined in the same way but with $k_2$ instead of $k_1$. $k_4$ also repeats this, but in this instance, $\frac{h}{2}$ is replaced with $h$. This gives us the following two equations:
\begin{equation}
    k_3=(x_n+\frac{h}{2}k_{2(x)},y_n+\frac{h}{2}k_{2(y)},\dot{x}_n+\frac{h}{2}k_{2(\dot{x})},\dot{y}_n+\frac{h}{2}k_{2(\dot{y})} )
\end{equation}
\begin{equation}
    k_4=(x_n+hk_{3(x)},y_n+hk_{3(y)},\dot{x}_n+hk_{3(\dot{x})},\dot{y}_n+hk_{3(\dot{y})} )
\end{equation}

Combining all of the above into a final weighted average gives us the final form of RK4:
\begin{equation}
    (x_{n+1},y_{n+1}, \dot{x}_{n+1},\dot{y}_{n+1}) = (x_n,y_n,\dot{x}_n,\dot{y}_n)+\frac{h}{6}(k_1+2k_2+2k_3+k_4)
\end{equation}
Lastly, this process is iterated over a total time of $T$ (see Section \ref{subsubsec:rk4}) with time steps $h$, where each $h$ represents an iteration of the RK4 method. RK4 was selected because of its fourth-order accuracy while remaining efficient for systems with rapidly varying accelerations, such as the CR3BP.

\section{Target Jacobi Constant}
\label{appendix:targetjacobiconstant}
Rather than manually specifying an initial velocity, it's possible to set an expected value for $C$, and then calculate the $|v|$ with that target $C$. From the equation for $C$ from Section \ref{subsec:zerovelocityfunction}, we get:
\begin{equation}
    \dot{x}^2+\dot{y}^2=2\Omega(x,y)-C_\text{target}
\end{equation}
$\dot{x}^2+\dot{y}^2$ can be written as $|v|^2$, where $|v|$ denotes the total velocity magnitude, simplifying the final equation to the following:
\begin{equation}
    |v|=\sqrt{2\Omega(x,y)-C_\text{target}}
\end{equation}
    
\end{document}